# A symplectic approach to gravitational instability

J. Perez and M. Lachièze-Rey

*CEA DSM/DAPNIA, Service d'Astrophysique, CE-Saclay,F-91191 Gif sur Yvette Cedex, France*

We present a global approach of non-dissipative physics. Based on symplectic mechanics this technique allows us to obtain the solution of a very large class of problems in terms of a Taylor expand. We apply this method to the problem of gravitational instability and we obtain a general expression of the gravitational potential, solution of the Vlasov-Poisson system, as a function of time in the context of Newtonian dust cosmology.

## I. INTRODUCTION

The Hamiltonian formalism has been proved since a long time to be very general and efficient in classical dynamics. In the recent years, Hamiltonian techniques have also been applied efficiently to continuous systems. Here we give a general presentation of Hamiltonian formalism (reflecting the underlying symplectic structure) applied to systems with a finite or infinite number of degrees of freedom. The latter case means systems described by a distribution function in phase space (like electro-magnetic or gravitational plasmas). Exploring evolution in time, i.e., motion, we are able to derive, for such systems, the analytical expression giving the evolution of any physical quantity as a function of initial conditions only. This solution is given in terms of a Taylor development, with a rule to calculate the coefficients.

Collisionless dynamics describes various astrophysical systems, typically self gravitating collapsing clouds of stars or dust particles. It may be applied, for instance, to study the stability and evolution of globular clusters, the formation of various cosmic structures by gravitational instability. Here we apply our Hamiltonian formalism to the study of gravitational instability, i.e., the evolution of a cosmic fluid under the influence of its own weight. Although our calculations are performed in the Newtonian approximation, they apply perfectly to the evolution of perturbations to the dust Friedmann – Lemaître universe, assumed to describe the first stages of galaxies or large scale structure formation. For such systems we are able to derive the expression of any physical quantity as a development with respect to time, depending on initial conditions only. We illustrate by developing the value of the (self–consistent) gravitational potential at any point and at any time, as a function of initial conditions only.

The paper is organized as follow : a first part is devoted to the presentation of the Hamiltonian formalism. Considering first systems with a finite number of degrees of freedom, we derive the evolution equation for any quantity, and explicit its solution. In order to show the mechanism of our formalism, we illustrate it on the very simple example of the harmonic oscillator. Then we generalize the formalism to continuous systems, by using functionals and functional derivatives, which allows us in this case also to derive the equation evolution for any quantity, as well as its solution under the form of a development. This allows in principle to calculate any physical quantity as a function of initial conditions.

The second part specializes the problem to the gravitational instability. We first recall the dynamical equations describing it, and their solution in the cosmological context (Newtonian approximation to describe the dust Friedmann–Lemaître models and their perturbations). Then we explicitly write the Poisson–Vlasov equations which describe the evolution of the system in the phase space, and its link with the usual hydrodynamical approach. Applying the results of the first part, we express the time–derivatives of physical quantities : as expected we recover the Liouville equation for the distribution function itself, and the derivative of the potential. We then explicit the Taylor development for the gravitational potential, and provide a recurrence formula which allows its explicit calculation from initial conditions.

## II. THE SYMPLECTIC APPROACH OF NON DISSIPATIVE PHYSICS

This section presents a new and very general formulation of the non dissipative physical equations, applicable to many kinds of different problems. The unification proposed here is a generalization of several different works ( [1], [6]). We split the framework of non dissipative physic into two parts : systems with a finite number $N$ of degrees of freedom on one hand, and we detail the case $N = 1$; on the other hand, systems with an infinite number of degrees of freedom, i.e., statistical physics. We will show that the equations for these two domains are strictly equivalent and derive from the same least action principle. We will propose moreover, the solution of this equation.

## A. Systems with N degrees of freedom

We will consider the 3-dimensional case, corresponding to one particle in 3-dimensional space. The variables are the spatial position $\mathbf{q}$, and its conjugated quantity, the linear momentum $\mathbf{p}$, both 3-vectors. The generalization to more dimensions is straightforward, the variables becoming any arbitrary collections of conjugated vectors (for systems of $N$ particles for instance), or tensors.

The foundation of our analysis relies on the possibility to associate a generator, a Hamiltonian-like function $g(\mathbf{q}, \mathbf{p})$, to any kind of transformation undergone by the system. In the following we will be mainly interested by one peculiar kind of transformation, i.e., motion. But transformations can also correspond to the effect of a class of perturbations on a given equilibrium state, or represent a geometrical change applied to the system (Rotation, Translation, etc...). In this sense, the formalism introduced here allows more than the simple study of motion : the search for invariances, the stability studies and, more generally, the systemization of the study of dynamical systems.

To the transformation considered is associated a parameter $\lambda$ which allows to follow the transformation. When motion is the transformation considered, $g$ is the usual Hamiltonian representing the energy $E$ of the system, and $\lambda$ identifies with the time $t$ (to study instabilities, $g$ will represent the characteristics of the instability, and $\lambda$ the intensity of the displacement). Then one can always define the meta-action as the function

$$s(\mathbf{p}, \mathbf{q}) = \int \left( \mathbf{p} \cdot \frac{d\mathbf{q}}{d\lambda} - g \right) d\lambda, \tag{1}$$

where $\mathbf{q}$ and $\mathbf{p}$ are functions of $\lambda$. In the case of motion, this meta-action reduces to the usual action, solution of the classical Hamilton-Jacobi equation (see [1] for example). In this peculiar case the total energy $E$ of the system is conserved through the $t$-evolution. Similarly, our formulation requires that the generator $g$ does not depend explicitly on $\lambda$ (although it depends through $\mathbf{q}$ and $\mathbf{p}$) and is conserved through the transformation.

Before turning to the key propositions, we recall some a standard definition and notations. The (usual) Poisson brackets for 2 functions $x(\mathbf{p}, \mathbf{q})$ and $y(\mathbf{p}, \mathbf{q})$ is defined as

$$P_3[x, y] := \sum_{i=1}^{3} \left( \frac{\partial x}{\partial q_i} \cdot \frac{\partial y}{\partial p_i} - \frac{\partial x}{\partial p_i} \cdot \frac{\partial y}{\partial q_i} \right), \tag{2}$$

and we also define the new bracket

$$L_3[x, y] := \int P_3[x, y] \, d\lambda, \tag{3}$$

which has all properties of a standard Lie bracket.

We now state the fundamental proposition, whose demonstration follows from the calculations below : the evolution of $\mathbf{p}$ and $\mathbf{q}$, as a function of $\lambda$, is given by the least action principle

$$L_3[k, s] = \int P_3[k, s] \, d\lambda = 0. \tag{4}$$

for any function $k(\mathbf{q}, \mathbf{p})$. This relation was initially presented in the special case of motion by [6] and [7].

Since (4) is true for any function $k$, it implies in particular $\forall i$ , $\frac{\partial s}{\partial q_i} = 0$ and $\forall i$ , $\frac{\partial s}{\partial p_i} = 0$, corresponding to the two evolution equations for $\mathbf{q}$ and $\mathbf{p}$ :

- Firstly,

$$\forall i \quad \frac{\partial s}{\partial p_i} = 0 \iff \frac{\partial}{\partial p_i} \int \left( \mathbf{p} \cdot \frac{d\mathbf{q}}{d\lambda} - g \right) d\lambda = 0$$
$$\iff \int \left( \frac{dq_i}{d\lambda} + \mathbf{p} \cdot \frac{\partial}{\partial p_i} \frac{d\mathbf{q}}{d\lambda} - \frac{\partial g}{\partial p_i} \right) d\lambda = 0.$$

Since $\mathbf{q}$ and $\mathbf{p}$ are explicitly independent, and $\mathbf{q}.\mathbf{p}$ vanishes on the surface of the system[1] , integration by parts gives

---

[1]This assumption makes no problems in this finite dimensional case, more details are required for continuous systems, see section C and appendix A

$$\int \left( \frac{dq_i}{d\lambda} - \frac{\partial g}{\partial p_i} \right) d\lambda \;=\; 0,$$

which implies

$$\frac{dq_i}{d\lambda} \;=\; \frac{\partial g}{\partial p_i} \;=\; \frac{\partial q_i}{\partial q_i}\frac{\partial g}{\partial p_i} - \frac{\partial q_i}{\partial p_i}\frac{\partial g}{\partial q_i} \;\Longleftrightarrow\; \frac{d\mathbf{q}}{d\lambda} \;=\; P_3[\mathbf{q},g] \;, \qquad (5)$$

which is the evolution equation for $\mathbf{q}$ induced by the transformation $g$.

- Similarly,

$$\forall i \;\; \frac{\partial s}{\partial q_i} = 0 \;\Longleftrightarrow\; \frac{\partial}{\partial q_i}\int \left( \mathbf{p}.\frac{d\mathbf{q}}{d\lambda} - g \right) d\lambda \;=\; 0$$

$$\Longleftrightarrow\; \int \left( \mathbf{p}.\frac{\partial}{\partial q_i}\frac{d\mathbf{q}}{d\lambda} - \frac{\partial g}{\partial q_i} \right) d\lambda \;=\; 0.$$

And, after a new integration by parts, we get

$$\frac{dp_i}{d\lambda} \;=\; -\frac{\partial g}{\partial q_i} \;=\; \frac{\partial p_i}{\partial q_i}\frac{\partial g}{\partial p_i} - \frac{\partial p_i}{\partial p_i}\frac{\partial g}{\partial q_i} \;\Longleftrightarrow\; \frac{d\mathbf{p}}{d\lambda} \;=\; P_3[\mathbf{p},g], \qquad (6)$$

the evolution equation for $\mathbf{p}$ induced by the transformation $g$.

In the special case of motion, where $g = E$ and $\lambda = t$, (5) and (6) are Hamilton's equations.

### B. Evolution equation for any function

More generally, the same technique allows to write the evolution equation of any function $k(\mathbf{q},\mathbf{p})$. Starting from the usual differentiation formula

$$\frac{dk}{d\lambda} = \sum_{i=1}^{3}\left( \frac{\partial k}{\partial q_i}\frac{dq_i}{d\lambda} + \frac{\partial k}{\partial p_i}\frac{dp_i}{d\lambda} \right) \qquad (7)$$

using (5) and (6) some straightforward algebra gives the evolution equation for $k$ induced by the transformation $g$

$$\frac{dk}{d\lambda} \;=\; P_3[k,g]. \qquad (8)$$

This is a linear and first order differential equation. Its solution is known to be (when $g$ is explicitly independent of $\lambda$)

$$k(\mathbf{q},\mathbf{p}) = k(\mathbf{q},\mathbf{p})\mid_{\lambda_o} -\lambda\, P_3[g,k]\mid_{\lambda_o} +\frac{\lambda^2}{2}\, P_3[g,P_3[g,k]]\mid_{\lambda_o} +\cdots, \qquad (9)$$

which is usually abbreviated under the notation

$$k(\mathbf{q},\mathbf{p}) = e^{\lambda P_3[.,g]}\; k(\mathbf{q},\mathbf{p})\mid_{\lambda_o}. \qquad (10)$$

This formulation is similar to the usual Hamiltonian formalism, with the main additional result here that explicit solution for the motion (or, more generally, the transformation) is given as a function of time (or, more generally $\lambda$). This offers several advantages in comparison with the classic formulation.
Firstly, we can expand any function in terms of any parameter. Although the case of motion is specially interesting, we have more general results. For example, taking for $g$ the generator of a perturbation, and for $k$ the energy of the system, we get a general energy variational principle for every system. In the context of stellar dynamics, [2], [8], [3] and [4] have produced new stability results. Finally, one can use this method to produce an integral of the evolution equation. A very important improvement lies in the fact that this formalism can be extended to systems with an infinite number of degrees of freedom. Before turning to it, we use a very simple application to illustrate how it works.

To illustrate some of the mechanism of the technique, we will shortly analyze here the trivial case of the motion of an one dimensional harmonic oscillator, with $\lambda = t$ and $g = E = \frac{p^2}{2m} + \frac{1}{2}\omega^2 q^2$. We use (9) with $k = |\mathbf{q}| := q$, and the initial conditions $q(t=0) = q_o$ and $p(t=0) = p_o$, we have

$$q(t) == q(t_o) - t P_3[E,q]\mid_{t=t_o} + \frac{t^2}{2} P_3[E, P_3[E,q]]\mid_{t=t_o} - \frac{t^3}{3!} P_3[E, P_3[E, P_3[E,q]]]\mid_{t=t_o} + \cdots . \tag{11}$$

The Poisson brackets are easy to compute :

$$P_3[E,q] = \frac{\partial E}{\partial q}\frac{\partial q}{\partial p} - \frac{\partial E}{\partial p}\frac{\partial q}{\partial q} = -\frac{p}{m}$$

$$P_3[E, P_3[E,q]] = P_3[E, -\frac{p}{m}] = -\frac{q\omega^2}{m}$$

$$P_3[E, P_3[E, P_3[E,q]]] = P_3[E, -\frac{q\omega^2}{m}] = \frac{p\omega^2}{m^2}$$

and so on. We can obtain a recurrence rule. Injecting these results in (11) one can have

$$\begin{aligned}
q(t) &= q_o + \frac{p_o}{m}t - \frac{1}{2}\frac{q_o\omega^2}{m}t^2 - \frac{t^3}{3!}\frac{p_o\omega^2}{m^2} - \frac{t^4}{4!}\frac{q_o\omega^4}{m^2} + \frac{t^5}{5!}\frac{p_o\omega^4}{m^2} + \cdots \\
&= q_o\left(1 - \frac{(\theta t)^2}{2!} + \frac{(\theta t)^4}{4!} + \cdots\right) + \frac{p_o}{m\theta}\left(\theta t - \frac{(\theta t)^3}{3!} + \frac{(\theta t)^5}{5!} + \cdots\right) \\
&= q_o \cos\theta t + \frac{p_o}{m\theta}\sin\theta t \qquad \text{with } \theta = \sqrt{\frac{\omega^2}{m}}
\end{aligned} \tag{12}$$

which is fortunately the well known motion equation of a free harmonic oscillator.

### C. Systems with an infinity of freedom's degree

The previous analysis was explicitly written for one particle ($N = 1$) but can be generalized to any $N$ in a straightforward way. Let us study now the $N \longrightarrow \infty$ limit, i.e., non–local statistical physics where variables are no more discrete. In this continuous limit, a dynamical system is no more described by a finite collection of $\{\mathbf{q}_i(\lambda), \mathbf{p}_i(\lambda)\}$ but through a distribution function $f(\mathbf{q}, \mathbf{p}, \lambda)$ in the phase space. We deal with a non dissipative problem, hence, we suppose that the system evolve without collisions between its components. We will show that the structure of the equations and the solution of the problem of evolution of such systems are exactly the same than in the finite case.

First, let us introduce a (non–relativistic) 4–dimensional formalism adapted to this study. We consider the 4–vectors $\mathbf{Q} = (\mathbf{q}, \lambda)$ and $\mathbf{P} = (\mathbf{p}, g)$, where $g$ and $\lambda$ are the independents quantities defined in the previous section, which generalize energy and time for any kind of transformation. In the following we will consider functionals of $f$, considered as a function of $\{\mathbf{q}, \mathbf{p}, \lambda\}$. In the space of functions of $\{\mathbf{q}, \mathbf{p}, \lambda\}$, we denote the functional derivative $\delta^+$. This notation is introduced to distinguish from the notation $\delta$ for the functional derivative in the space of functions of $\{\mathbf{q}, \mathbf{p}\}$ only, to be used later. We now define the meta–action $\mathcal{S}[f]$ through the relation $\frac{\delta^+ \mathcal{S}}{\delta^+ f} = g$. In the special case where $g$ does not depend on $f$, it can be written as

$$\mathcal{S}[f] := \int f.g\, d^3\mathbf{q}\, d^3\mathbf{p}\, d\lambda, \tag{13}$$

but this linear formula does not apply, for instance, to the gravitational instability problem. We recall that we have assumed the generator $g$ independent of $\lambda$ and we note that the meta–action $\mathcal{S}$ does not depend either on $\lambda$. We also introduce another functional $\mathcal{G}$, now defined over the space of functions of $\{\mathbf{q}, \mathbf{p}\}$ only (thus at given $\lambda$), and such that $\frac{\delta \mathcal{G}}{\delta f} = g$. For the linear case where $g$ does not depend on $f$, it can be written

$$\mathcal{G}[f] := \int f.g. \, d^3\mathbf{q} \, d^3\mathbf{p}. \tag{14}$$

Note that, in general, $\mathcal{G}$ depends on $\lambda$. Let us define now a 4-dimensional Poisson bracket in the space of functions of $\{\mathbf{Q}, \mathbf{P}\}$ by

$$\forall \; \mathcal{X}, \mathcal{Y} \quad P_4[\mathcal{X}, \mathcal{Y}] := \sum_{i=1}^{4} \frac{\partial \mathcal{X}}{\partial Q_i} \frac{\partial \mathcal{Y}}{\partial P_i} - \frac{\partial \mathcal{X}}{\partial P_i} \frac{\partial \mathcal{Y}}{\partial P_i}, \tag{15}$$

and a 4-dimensional Lie-bracket, relative to the distribution function $f$, in the space of functionals, as

$$\forall \; \mathcal{X}, \mathcal{Y} \quad L_4[\mathcal{X}, \mathcal{Y}]_{(f)} := \int f.P_4[\frac{\delta^+ \mathcal{X}}{\delta^+ f}, \frac{\delta^+ \mathcal{Y}}{\delta^+ f}] \tag{16}$$

We are now able to show that the evolution equation of the system is equivalent to

$$\forall \, \mathcal{K} \quad L_4[\mathcal{K}, \mathcal{S}]_{(f)} = 0. \tag{17}$$

Taking into account $\delta^+ \mathcal{S}[f]/\delta^+ f = g$, the latter relation is equivalent to

$$\forall \, \mathcal{K} \quad \int f.P_4[\frac{\delta^+ \mathcal{K}}{\delta^+ f}, g] \, d^3\mathbf{q} \, d^3\mathbf{p} \, d\lambda = 0. \tag{18}$$

Following [7], an integration by parts (see appendix A) then gives

$$\forall \, \mathcal{K} \quad \int \frac{\delta^+ \mathcal{K}}{\delta^+ f}.P_4[g, f] \, d^3\mathbf{q} \, d^3\mathbf{p} \, d\lambda = 0. \tag{19}$$

Since $\mathcal{K}$ can be any functional, this last relation reduces to $\mathcal{P}_4[g, f] = 0$. Developing

$$P_4[g, f] = P_3[g, f] + \frac{\partial g}{\partial \lambda}\frac{\partial f}{\partial g} - \frac{\partial g}{\partial g}\frac{\partial f}{\partial \lambda} = 0 \tag{20}$$

and taking into account that $g$ is explicitly independent of $\lambda$, the equation is finally reduced to

$$\frac{\partial f}{\partial \lambda} = P_3[g, f]. \tag{21}$$

This equation describes the evolution of the distribution function $f$ of a non dissipative system. For the special case of motion, it is easy to check that it reduces to the usual Liouville equation. Making the mean field approximation, we recover the Vlasov dynamicalequation of collisionless systems. For continuous systems, observables are obtained by averaging physical quantities following the distribution function in phase space, thus observables may be seen as functionals $\mathcal{K}$ acting on a positive and normalisable function $f$ of $\{\mathbf{q}, \mathbf{p}, \lambda\}$. Moreover for any such functional we have

$$\forall \mathcal{K} \quad \frac{d\mathcal{K}[f]}{d\lambda} := \int \frac{\delta \mathcal{K}}{\delta f} \frac{\partial f}{\partial \lambda} \, d^3\mathbf{q} \, d^3\mathbf{p}. \tag{22}$$

Injecting (21) one has

$$\frac{d\mathcal{K}}{d\lambda} = \int \frac{\delta \mathcal{K}}{\delta f} P_3[g, f] \, d^3\mathbf{q} \, d^3\mathbf{p}. \tag{23}$$

Using $g = \frac{\delta \mathcal{G}}{\delta f}$, an integration by part (see appendix A) then gives

$$\frac{d\mathcal{K}}{d\lambda} = \int f \, P_3[\frac{\delta \mathcal{K}}{\delta f}, \frac{\delta \mathcal{G}}{\delta f}] \, d^3\mathbf{q} := M_3[\mathcal{K}, \mathcal{G}]_{(f)} \tag{24}$$

where we have defined the new Lie brackets $M_3[\mathcal{K}, \mathcal{G}]_{(f)}$ relative to the distribution function $f$. As for the discrete systems, $M_3[., \mathcal{G}]$ is an $\lambda$-independent operator, thus (24) can be solved using the Taylor expand of the exponential

$$\mathcal{K}[f] = \mathcal{K}[f_o] - \lambda \, M_3[\mathcal{G}, \mathcal{K}]_{(f)} |_{\lambda = \lambda_o} + \frac{\lambda^2}{2} M_3[\mathcal{G}, M_3[\mathcal{G}, \mathcal{K}]]_{(f)} |_{\lambda = \lambda_o} + \cdots \tag{25}$$

For dynamical evolution (motion), the generator functional $\mathcal{G}$ is the Hamiltonian $\mathcal{H}[f]$ of the system. Thus the time–evolution of any functional $\mathcal{K}$ is given by

$$\mathcal{K}[f] = \mathcal{K}[f_o] - t\, M_3[\mathcal{H}, \mathcal{K}]_{(f)}\,|_{t=t_o} + \frac{t^2}{2} M_3[\mathcal{H}, M_3[\mathcal{H}, \mathcal{K}]]_{(f)}\,|_{t=t_o} + \cdots \tag{26}$$

This latter expansion gives the expression of any quantity (mean potential, entropy ... ) of the system, at time $t$, as a function of initial conditions only. This may be applied to many problems, as far as the brackets are known. In the following, we apply it to the problem of gravitational instability.

All these results can be resumed in a little box, expressing the symmetries between discrete and continuous non dissipative systems

---

**One degree of freedom**

Variables : $\mathbf{q}$, $\mathbf{p}$, $\lambda$

Meta-Action : $s(\mathbf{q}, \mathbf{p}) := \int (\mathbf{p} \frac{d\mathbf{q}}{d\lambda} - g) d\lambda$

Evolution $\begin{cases} \forall k(\mathbf{q},\mathbf{p})\ L_3[k,s](\mathbf{q},\mathbf{p}) = 0 \\ \quad \Longleftrightarrow \\ \frac{dk}{d\lambda} = P_3[k,g] \end{cases}$

Solution : $k(\mathbf{q},\mathbf{p}) = e^{\lambda P_3[.,g]} k(\mathbf{q},\mathbf{p})|_{\lambda=\lambda_o}$

---

**Infinity of freedom's degrees**

Variable : $f(\mathbf{q}, \mathbf{p}, \lambda)$

Meta-Action : $\mathcal{S}[f] := \int f.g.d^3\mathbf{q} d^3\mathbf{p} d\lambda$

Evolution $\begin{cases} \forall\, \mathcal{K}[f]\ L_4[\mathcal{F}, \mathcal{S}]_{(f)} = 0 \\ \quad \Longleftrightarrow \\ \frac{d\mathcal{K}}{d\lambda} = M_3[\mathcal{K}, \mathcal{G}] \end{cases}$

Solution : $\mathcal{K}[f] = e^{\lambda M_3[.,\mathcal{G}]} \mathcal{K}[f]|_{\lambda=\lambda_o}$

---

## III. GRAVITATIONAL INSTABILITY

### A. Dynamical equations

Here we apply the previous results to the problem of Gravitational instability. This problem was introduced in this context by [4]. We will however not consider in this paper the standard formulation of the problem, using perturbed quantities, with respect to the Friedmann–Lemaître, and a comoving formulation. This formulation would imply explicit time dependences in the evolution equations or the Hamiltonian, which would forbid the application of the previous rules. We rather calculate the evolution of the total (unperturbed plus perturbed) quantities. Since the evolution of the unperturbed ones are trivially known, under the form of the cosmological models, a simple difference will provide the perturbations. We will consider the universe as filled of dust, i.e., pressureless matter only, and apply the Newtonian approximation as usual. Since gravitational instability becomes non linear much after the period of matter-radiation equivalence, i.e., when Universe is dominated by pressureless matter, this approximation applies well.

The evolution of cosmic matter is described by the dynamical equation

$$\frac{d^2 \mathbf{q}}{dt^2} = \mathbf{g} = -\nabla_{\mathbf{q}} \phi, \tag{27}$$

and the Poisson equation

$$\Delta_{\mathbf{q}} \phi = 4\pi G\ \rho, \tag{28}$$

where $\rho$ is the matter density and $\phi$ the gravitational potential. The velocity $\mathbf{V} = \frac{d\mathbf{q}}{dt}$. The time derivatives in these equations are to be taken following the motion, so that $\frac{d}{dt} = \frac{\partial}{\partial t} + \mathbf{V}.\nabla_{\mathbf{q}}$. Hereafter, primes will denote partial derivatives with respect to time $\frac{\partial}{\partial t}$, i.e., derivative at a fixed (Eulerian) position $\mathbf{q}$. The density evolves according to mass conservation $\rho' = -div(\rho\ \mathbf{V})$.

It is well known that it is possible to express the usual Friedmann–Lemaître models in a (Newtonian) form obeying these equations. This will be our unperturbed solution :

$$\rho_u = \rho_0\ a^{-3} \tag{29}$$
$$\mathbf{V}_u = H\ \mathbf{q} = (a'/a)\ \mathbf{q} \tag{30}$$
$$\phi_u = 2\pi G\ \rho_u\ \mathbf{q}^2/3, \tag{31}$$

where the scale factor $a(t)$ obeys the Friedmann equations

$$a^2\ a'' = -\gamma/2 \tag{32}$$

and

$$a'^2 = \gamma a^{-1} - k, \tag{33}$$

where we have defined $\gamma \equiv 8\pi\ G\ \rho_0/3$. The scale factor $a$ is normalized to its value at a time $t_0$, chosen as an origin and $\rho_0$ denotes the value of the density at $t_0$ ; $k = -1, 0$, or $1$ is the curvature factor of space.

### B. The Vlasov–Poisson system

Gravitational instability is usually described by hydrodynamic equations, to which the Poisson equation is added to express self–gravity. However, it is not always possible to find a Hamiltonian formulation of hydrodynamics. It is thus advantageous to work with the Vlasov–Poisson system, which is more general, and from which hydrodynamics can be deduced as a peculiar case.

Thus we will consider equation (27) as the dynamical part of a Vlasov equation. This latter, which allows to follow the velocity distribution, is advantageous to consider for the Hamiltonian formulation. Hydrodynamics will be treated as a peculiar case of the Vlasov–Poisson formulation. Assuming particles with unit mass, the matter is described by a distribution function $f(\mathbf{q}, \mathbf{p}, t)$ such that

$$\int d^3\mathbf{p}\ \mathbf{p} f(\mathbf{q}, \mathbf{p}, t), \tag{34}$$

and the average impulsion at point $\mathbf{q}$

$$\rho(\mathbf{q}, t)\ \mathbf{V}\ =\ \int d^3 p\ \mathbf{p}\ f(\mathbf{q}, \mathbf{p}, t)\ . \tag{35}$$

Note that the unperturbed distribution function takes the simple form $f_u(\mathbf{q},\ \mathbf{p}, t) = \rho_0\ a(t)^{-3}\ \delta_D(\mathbf{p} - H(t)\mathbf{q})$, where $\delta_D$ is the Dirac function. With these definitions, the Liouville equation takes the form

$$\frac{\partial f}{\partial t} + \mathbf{p}.\frac{\partial f}{\partial \mathbf{q}} - \frac{d\phi}{d\mathbf{q}}.\frac{\partial f}{\partial \mathbf{p}} = 0. \tag{36}$$

As we will check, the kinetic and potential "energies" associated to one particle are respectively $T = \mathbf{p}^2/2$ and $\phi(\mathbf{q})$. The gravitational potential at a point $\mathbf{q}_0$ is obtained by integrating (28) as

$$\phi(\mathbf{q}_0, t) = -G \int d\Gamma \, \frac{f(\mathbf{q}, \mathbf{p}, t)}{|\mathbf{q} - \mathbf{q}_0|}, \tag{37}$$

where $d\Gamma = d^3\mathbf{p} d^3\mathbf{p}$ is the phase space volume element.

### C. Hamiltonian formulation of gravitational instability

#### 1. The Vlasov–Poisson system

Considering gravitational instability as a motion with respect to time, its generator is the Hamiltonian $g = T(\mathbf{p}) + \phi$, corresponding to the functional integral

$$\mathcal{G} = \int d\Gamma \, G/2 \int d\Gamma \int d\Gamma' \frac{f(\mathbf{q}, \mathbf{p}, t) \, f(\mathbf{q}', \mathbf{p}', t)}{|\mathbf{q} - \mathbf{q}_0|} \tag{38}$$

so that the functional derivative $\frac{\delta \mathcal{G}}{\delta f} = g$. The application of the results of the previous sections involves the brackets $M_3[\mathcal{A}, \mathcal{G}]$ (with respect to the function $f$). In order to calculate them, we need to use the following relations, straightforward to derive :

$$\frac{\partial}{\partial \mathbf{p}} \left( \frac{\delta \mathcal{G}}{\delta f} \right) = \frac{\partial g}{\partial \mathbf{p}} = \mathbf{p} \tag{39}$$

and

$$\frac{\partial}{\partial \mathbf{q}} \left( \frac{\delta \mathcal{G}}{\delta f} \right) = \frac{\partial g}{\partial \mathbf{q}} = \phi_{,\mathbf{q}}, \tag{40}$$

where we wrote $\phi_{,\mathbf{q}} = \frac{\partial \phi}{\partial \mathbf{q}}$.

This allows to calculate the evolution of any functional $\mathcal{A}$ through the fundamental relation

$$\frac{d\mathcal{A}}{dt} = M_3[\mathcal{A}, \mathcal{G}]. \tag{41}$$

Let us consider a functional $\mathcal{A}$ with its functional derivative $\frac{\delta \mathcal{A}}{\delta f} = A$. When $A$ does not depends on $f$ (linear case), equation (41) implies that

$$\frac{d\mathcal{A}}{dt} = \int d\Gamma \, A^1(\mathbf{q}, \mathbf{p}, t) \, f(\mathbf{q}, \mathbf{p}, t), \tag{42}$$

where

$$A^1 = P_3[A, g] = \frac{\partial A}{\partial \mathbf{q}} \cdot \frac{\partial g}{\partial \mathbf{p}} - \frac{\partial A}{\partial \mathbf{p}} \cdot \frac{\partial g}{\partial \mathbf{q}} = \mathbf{p} \cdot \frac{\partial A}{\partial \mathbf{q}} - \phi_{,\mathbf{q}} \cdot \frac{\partial A}{\partial \mathbf{p}}. \tag{43}$$

On the other hand, when $A$ depends on $f$ (non linear case), the situation is more complicated and, as we will see later, the functional derivative of $A$ must appear in $\frac{d\mathcal{A}}{dt}$.

#### 2. Hydrodynamics

It is often sufficient – and this is the general approach – to consider gravitational instability from an hydrodynamic rather than Vlasov point of view. However there is no general Hamiltonian treatment for hydrodynamics (except in the case of irrotational flow). To recover hydrodynamics, we may start from the Vlasov description given above, and take the appropriate moments, neglecting the convenient terms, in the standard manner. In fact it is appropriate to remark that hydrodynamics for cold matter can be recovered by specifying the distribution function as $f(\mathbf{q}, \mathbf{p}, t) =$ $\rho(\mathbf{q},t) \, \delta_D(\mathbf{p} - \mathbf{V}_{(\mathbf{q},t)})$. Since the gas is considered to be cold and pressureless, all values of $\mathbf{p}$ at point $\mathbf{q}$ are equal to $\mathbf{V}$. This allows to recover exactly the usual description of gravitational instability.

The unperturbed distribution function for hydrodynamics is thus

$$f_u(\mathbf{q},\mathbf{p},t) = \rho_u(\mathbf{q},t) \, \delta_D(\mathbf{p} - \mathbf{V}_{u(\mathbf{q},t)}) \tag{44}$$

In the hydrodynamical case, the derivation rule derived above takes a simple form. When we are interested in the evolution of a functional quantity of the type

$$\mathcal{A} = \int d^3\mathbf{q} \, \rho(\mathbf{q},t) A(\mathbf{q}, \mathbf{V}) \, ,$$

the application of the previous results leads directly to the derivation formula

$$\mathcal{A}' = \int d^3\mathbf{q} \, \rho(\mathbf{q},t) A(\mathbf{q}, \mathbf{V}) \, ,$$

where $A$ and $A^{(1)}$ are related by (43).

### D. Time derivatives

#### 1. Time derivative of the distribution function

It may be more interesting to evaluate the derivative of the gravitational potential at a point $\mathbf{q}_0$. The expression (37) defines $\phi(\mathbf{q}_0,t)$ as a functional of $f$, such that

$$A(q) = \frac{\delta \phi(\mathbf{q}_0,t)}{\delta f} = \frac{-G}{|\mathbf{q} - \mathbf{q}_0|}. \tag{45}$$

Clearly,

$$\frac{d}{d\mathbf{p}}\left(\frac{\delta \phi}{\delta f}\right) = 0 \tag{46}$$

and

$$\frac{d}{d\mathbf{q}}\left(\frac{\delta \phi}{\delta f}\right) = -G \, \frac{d}{d\mathbf{q}}\left(\frac{1}{|\mathbf{q} - \mathbf{q}_0|}\right). \tag{47}$$

Applying the results of previous sections, we may now calculate the time derivative of $\phi$ at point $\mathbf{q}_0$ :

$$\phi'(\mathbf{q}_0,t) = -G \int d^3\mathbf{q} \, d^3\mathbf{p} \, f(\mathbf{q},\mathbf{p},t) \, u_1(\mathbf{q},\mathbf{p};\mathbf{q}_0), \tag{48}$$

where

$$u_1(\mathbf{q},\mathbf{p};\mathbf{q}_0) = -\frac{1}{G} P_3 \left[\frac{\delta \phi}{\delta f}, \frac{\delta \mathcal{G}}{\delta f}\right] = \mathbf{p} \cdot \frac{d}{d\mathbf{q}}\left(\frac{1}{|\mathbf{q} - \mathbf{q}_0|}\right).$$

Integration by parts, in the hydrodynamical approximation, leads to

$$\phi'(\mathbf{q}_0,t) = G \int d^3q \, \frac{1}{|\mathbf{q} - \mathbf{q}_0|} \, div(\rho \, \mathbf{V}). \tag{49}$$

This equation could have been found directly by deriving (37) with respect to time and using the mass conservation $\rho' = -div(\rho \, \mathbf{V})$ but here we illustrate how the method works to calculate any time–derivative.

## 2. Taylor development

We have calculated the time–derivatives of two peculiar functionals, the distribution function itself and the potential. But we can also calculate, by applying iteratively the method, the derivatives at any order, of any functional $\mathcal{A}$. This will allow, for instance, to calculate its Taylor development. For any functional $\mathcal{A}$ with functional derivative $A$, we have shown that the time derivative $\mathcal{A}'$ may be written as the integral of $A^{(1)}$ related to $A$ by (43). It results that the n$^{th}$ order time derivative of $\mathcal{A}$ is of the form

$$\mathcal{A}^{(n)} = \int d\Gamma \ f \ A^{(n)} \tag{50}$$

where the $A^{(n)}$ can be calculated from the $A^{(n-1)}$. This calculation involves the functional derivative

$$\left(\frac{\delta \mathcal{A}^{(n)}}{\delta f}\right)(\mathbf{q},\mathbf{p}) = A^{(n)}(\mathbf{q},\mathbf{p}) + \int d\Gamma' \ f' \ \left(\frac{\delta A^{(n)}(\mathbf{q'},\mathbf{p'})}{\delta f}\right)(\mathbf{q},\mathbf{p}). \tag{51}$$

In this relation the last integral term takes into account the non linear case in which $A^{(n)}$ depends on $f$. From the above calculations, it results that

$$A^{(n+1)}(\mathbf{q},\mathbf{p}) = (\mathbf{p}.\frac{d}{d\mathbf{q}} - \phi_\mathbf{q}.\frac{d}{d\mathbf{p}}) \left[A^{(n)}(\mathbf{q},\mathbf{p}) + \int d\Gamma' \ f' \ \left(\frac{\delta A^{(n)}(\mathbf{q'},\mathbf{p'})}{\delta f}\right)(\mathbf{q},\mathbf{p})\right]. \tag{52}$$

which reduces to only

$$A^{(n+1)}(\mathbf{q},\mathbf{p}) = (\mathbf{p}.\frac{d}{d\mathbf{q}} - \phi_\mathbf{q}.\frac{d}{d\mathbf{p}}) \ [A^{(n)}(\mathbf{q},\mathbf{p})], \tag{53}$$

in the linear case where $A^{(n)}$ does not depend on $f$.

Let us apply that to calculate some higher order derivatives of the potential. Let us first introduce the notation $J(\mathbf{q} - \mathbf{q}_0) := \frac{1}{|\mathbf{q}-\mathbf{q}_0|}$. Also, we will note the spatial derivatives any quantity $\Lambda$ as

$$\Lambda_{,ij\ldots k} \equiv d_{q_i} \ d_{q_j} \ldots d_{q_k} \ \Lambda. \tag{54}$$

For instance, the $J_{,i} = -\frac{(\mathbf{q}-\mathbf{q}_0)_i}{|\mathbf{q}-\mathbf{q}_0|^3}$ are the components of a vector. With more indices, $J_{,ij\ldots k}$ is a tensor. From $\phi'_{(\mathbf{q}_0,t)} = G \int d\Gamma \ f(\mathbf{q},\mathbf{p},t) \ [-p_i \ J_{,i}(\mathbf{q} - \mathbf{q}_0)]$, we can apply the previous formula to derivate again as

$$\phi''_{(\mathbf{q}_0,t)} = G \int d\Gamma \ f(\mathbf{q},\mathbf{p},t) \ u_2(\mathbf{q},\mathbf{p};\mathbf{q}_0), \tag{55}$$

with

$$u_2 = -\left(p_j \frac{d}{d_{q_j}} - \phi_{,j}\frac{d}{d_{p_j}}\right) \ p_i \ J_{,i}(\mathbf{q} - \mathbf{q}_0) = -p_i \ p_j \ J_{,ij}(\mathbf{q} - \mathbf{q}_0) + \phi_{,i} \ J_{,i}(\mathbf{q} - \mathbf{q}_0). \tag{56}$$

Reporting in the above formula, integrating by parts, and using the hydrodynamical approximation, we obtain

$$\phi''_{(\mathbf{q}_0,t)} = -G \int d^3\mathbf{q} \ \frac{1}{|\ \mathbf{q} - \mathbf{q}_0\ |} \left(\frac{d^2[\rho \ V_i \ V_j]}{dq_i \ dq_j} + \frac{d[\rho \ \frac{d\phi}{dq_i}]}{dq_i}\right). \tag{57}$$

Subsequent derivation leads to higher order terms. Care must be taken however that the functional derivative of $\phi''$ is not $u_2$ which appears in the previous formula, since $u_2$ is non linear and depends on the potential $\phi$ which depends itself on $f$. Thus, to calculate $\phi^{(3)}$ from $\phi^{(2)} \equiv \phi''$, we must come back to original formulae (51). First we calculate the functional derivative

$$\left(\frac{\delta \phi''_{(\mathbf{q}_0,t)}}{\delta f}\right)(\mathbf{q},\mathbf{p}) = u_2(\mathbf{q},\mathbf{p};\mathbf{q}_0) + \int d\Gamma' \ f' \ \left(\frac{\delta u_2(\mathbf{q'},\mathbf{p'};\mathbf{q}_0)}{\delta f}\right)(\mathbf{q},\mathbf{p}).$$

We continue by calculating the functional derivative of $u_2$ as :

$$\left(\frac{\delta u_2(\mathbf{q'},\mathbf{p'},\mathbf{q}_0)}{\delta f}\right)(\mathbf{q},\mathbf{p}) = J_{,i}(\mathbf{q'} - \mathbf{q}_0)\left(\frac{\delta \phi'_i(\mathbf{q'})}{\delta f}\right)(\mathbf{q}).$$

After some calculations with no difficulty, the integral in the previous formula may be written

$$G \int d^3\mathbf{q'} \; \rho(\mathbf{q'}) \; \frac{(\mathbf{q'} - \mathbf{q}_0).(\mathbf{q} - \mathbf{q'})}{\mid \mathbf{q'} - \mathbf{q}_0 \mid^3 \mid \mathbf{q} - \mathbf{q'} \mid^3}.$$

Finally, it results that

$$\phi^{(3)}_{(\mathbf{q}_0,t)} = G \int \frac{d^3\mathbf{q}}{\mid \mathbf{q} - \mathbf{q}_0 \mid} \left( \frac{d^3[\rho \; V_i \; V_j \; V_k]}{dq_i \; dq_j \; dq_k} + 3 \frac{d^2[\rho \; V_i \frac{d\phi}{dq_j}]}{dq_i \; dq_j} - \frac{d[\rho \; V_i \frac{d^2\phi}{dq_i \; dq_j}]}{dq_j} \right) \tag{58}$$

$$+ G \int d^3\mathbf{q'} \; \rho(\mathbf{q'}) \; \frac{(\mathbf{q'} - \mathbf{q}_0).(\mathbf{q} - \mathbf{q'})}{\mid \mathbf{q'} - \mathbf{q}_0 \mid^3 \mid \mathbf{q} - \mathbf{q'} \mid^3}. \tag{59}$$

Similar calculations would allow to calculate the next order terms and the Taylor development of the potential may be written as

$$\phi(\mathbf{q}_0,t) = \phi(\mathbf{q}_0,t=0)$$
$$+ G \int d\Gamma \; f_0(\mathbf{q'},\mathbf{p'}) \; (t \; u_1 \; + \; t^2 \frac{u_2}{2} \; + \; t^3 \frac{u_3}{6} + \ldots), \tag{60}$$

where all quantities are estimated at $t = 0$.

There is no peculiar difficulty with the resulting formulae, at any order, excepted for their length, so that we will not write them explicitly. The potential at any point, and at any time (even after shell–crossing), may be calculated as a function of initial conditions only. But the expression is non local, involving the integration of initial quantities (density, velocity and potential) over the whole space. As expected, the expressions found are non linear, and it is easy to check that their degree increases with the degree in the time development. Thus the formulae here are comparable to the perturbative developments usually performed. Also it appears that each further degree in the development involves high order spatial derivatives of the initial physical quantities, integrated over space. This suggests that, the smoother the initial condition, the closer the solution remains from the linear one.

## IV. CONCLUSION

We have presented a very general method to deal with Hamiltonian systems. Although the illustrating case is motion, i.e., evolution in time, this applies to any kind of evolution which can be described as a function of a time–like parameter, like invariances, instabilities, optics *etc.*.. For such systems, the evolution is described by a set of canonical variable (position and momentum for discrete systems), or by a distribution function in a continuous approach.
Having introduced operators which express the algebraic structure of the problem, we have been able to present an evolution equation for any quantity depending on the canonical variables, or on the distribution function. We have also presented a formal solution of this equation, which allows to calculate any physical quantity as a function of "initial" conditions only, although under the form of a development with an infinite number of terms.

These results apply well to the self–consistent problem of gravitational instability, where the distribution function evolves in time under its own gravitational interaction. This problem is treated here in the Newtonian approximation which applies in fact in the cosmological context, when fluctuations evolve with respect to a (dust) Friedmann–Lemaître model. We present formulae which give an account of the evolution of any quantity depending on the distribution function (or on the mass density and velocity field in the hydrodynamic approximation). In particular we calculate the self–consistent evolution of the gravitational potential. We have expressed its value at any point, and at any time, under the form of a Taylor development whose terms of any order may be calculated as functions of the initial conditions only. This apply to the whole (unperturbed + perturbed) potential but it is easy to subtract the unperturbed part, which is known analytically, to derive the evolution of the perturbation. This will offer a practical way to calculate the evolution of the potential (or of any quantity like the density perturbations or the velocity fields), in the context of the gravitational instability theory for the formation of galaxies and large scale structures.

In further work, we will compare the development introduced here with other development introduced in perturbative (Eulerian or Lagrangian) approaches. We will also treat in more details a case usually considered in gravitational instability scenarii, that with no initial velocity perturbations. In any case, this approach offers a new way to attack the question of gravitational instability.

## ACKNOWLEDGMENTS

J. Perez thanks J. M. Alimi who has initialized our discussions.

# APPENDIX : INTEGRATION BY PARTS AND SURFACE TERMS

Let us consider 3 functions $x(\mathbf{q},\mathbf{p})$, $y(\mathbf{q},\mathbf{p})$ and $z(\mathbf{p},\mathbf{q})$. We have

$$\iint x P_3[y,z] d^3\mathbf{q}\, d^3\mathbf{p} \;=\; \iint x \left( \frac{\partial^3 y}{\partial \mathbf{q}^3} \frac{\partial^3 z}{\partial \mathbf{p}^3} - \frac{\partial^3 y}{\partial \mathbf{p}^3} \frac{\partial^3 z}{\partial \mathbf{q}^3} \right) \tag{A1}$$

each term of the RHS of (A1) can be integrated by parts,

$$\iint x \frac{\partial^3 y}{\partial \mathbf{q}^3} \frac{\partial^3 z}{\partial \mathbf{p}^3} d^3\mathbf{q}\, d^3\mathbf{p} = \int \left| x\, y\, \frac{\partial^3 z}{\partial \mathbf{p}^3} \right|_{S_\mathbf{p}} d^3\mathbf{q} - \iint y \left( \frac{\partial^3 x}{\partial \mathbf{p}^3} \frac{\partial^3 z}{\partial \mathbf{q}^3} + x \frac{\partial^6 z}{\partial \mathbf{p}^3 \partial \mathbf{q}^3} \right) d^3\mathbf{q}\, d^3\mathbf{p} \tag{A2}$$

for the first term and

$$\iint x \frac{\partial^3 y}{\partial \mathbf{p}^3} \frac{\partial^3 z}{\partial \mathbf{q}^3} d^3\mathbf{q}\, d^3\mathbf{p} = \int \left| x\, y\, \frac{\partial^3 z}{\partial \mathbf{q}^3} \right|_{S_\mathbf{q}} d^3\mathbf{p} - \iint y \left( \frac{\partial^3 x}{\partial \mathbf{q}^3} \frac{\partial^3 z}{\partial \mathbf{p}^3} + x \frac{\partial^6 z}{\partial \mathbf{p}^3 \partial \mathbf{q}^3} \right) d^3\mathbf{q}\, d^3\mathbf{p} \tag{A3}$$

where $S_\mathbf{q}$ and $S_\mathbf{p}$ represents respectively the two 3-dimensional surfaces obtained when $\mathbf{q}$ and $\mathbf{p}$ goes to $\infty$. Injecting (A2) and (A3) into (A1) one can get

$$\iint x \mathcal{P}_3[y,z] d^3\mathbf{q}\, d^3\mathbf{p} = \iint y \mathcal{P}_3[z,x] d^3\mathbf{q}\, d^3\mathbf{p} + \int \left| x\, y\, \frac{\partial^3 z}{\partial \mathbf{p}^3} \right|_{S_\mathbf{p}} d^3\mathbf{q} - \int \left| x\, y\, \frac{\partial^3 z}{\partial \mathbf{q}^3} \right|_{S_\mathbf{q}} d^3\mathbf{p} \tag{A4}$$

In this last relation, the two last terms generally vanishes. Indeed, if

$$\lim_{\mathbf{p} \to \infty} x\, y\, \frac{\partial^3 z}{\partial \mathbf{p}^3} = 0 \quad \text{and} \quad \lim_{\mathbf{q} \to \infty} x\, y\, \frac{\partial^3 z}{\partial \mathbf{q}^3} = 0 \tag{A5}$$

then these surface terms vanishes. For example if, $x$ or $y$ is the distribution function in the phase space of a Newtonian finite system, both this terms vanishes. For the unperturbed solution describing the cosmological models, it can also be checked that all surface terms also vanish, so that integration by parts can be applied as well.

14